\documentclass[amsmath,amssymb,aps,prl,superscriptaddress,twocolumn,notitlepage,nofootinbib]{revtex4-1}
\usepackage[dvips]{graphicx}
\usepackage[USenglish]{babel}
\usepackage{amsmath,amssymb,amsthm,dsfont,url,braket,hyperref, mathtools,commath}
\usepackage[percent]{overpic}
\usepackage{xcolor}
\usepackage{float}

\begin{document}

\title{Device independent certification of a quantum delayed choice experiment}

\author{Emanuele Polino}
\affiliation{Dipartimento di Fisica, Sapienza Universit\`{a} di Roma,
Piazzale Aldo Moro 5, I-00185 Roma, Italy}

\author{Iris Agresti}
\affiliation{Dipartimento di Fisica, Sapienza Universit\`{a} di Roma,
Piazzale Aldo Moro 5, I-00185 Roma, Italy}

\author{Davide Poderini}
\affiliation{Dipartimento di Fisica, Sapienza Universit\`{a} di Roma,
Piazzale Aldo Moro 5, I-00185 Roma, Italy}

\author{Gonzalo Carvacho}
\affiliation{Dipartimento di Fisica, Sapienza Universit\`{a} di Roma,
Piazzale Aldo Moro 5, I-00185 Roma, Italy}

\author{Giorgio Milani}
\affiliation{Dipartimento di Fisica, Sapienza Universit\`{a} di Roma,
Piazzale Aldo Moro 5, I-00185 Roma, Italy}

\author{Gabriela Barreto Lemos}
\affiliation{International Institute of Physics, Universidade Federal do Rio Grande do Norte, Campus Universitario, Lagoa Nova, Natal-RN 59078-970, Brazil}

\author{Rafael Chaves}
\email{rchaves@iip.ufrn.br}
\affiliation{International Institute of Physics, Universidade Federal do Rio Grande do Norte, Campus Universitario, Lagoa Nova, Natal-RN 59078-970, Brazil}

\author{Fabio Sciarrino}
\email{fabio.sciarrino@uniroma1.it}
\affiliation{Dipartimento di Fisica, Sapienza Universit\`{a} di Roma,
Piazzale Aldo Moro 5, I-00185 Roma, Italy}

\begin{abstract}
Wave-particle duality has long been considered a fundamental signature of the non-classical behavior of quantum phenomena, specially in a delayed choice experiment (DCE), where the experimental setup revealing either the particle or wave nature of the system is decided after the system has entered the apparatus. However, as counter-intuitive as it might seem, usual DCEs do have a simple causal explanation. Here, we take a different route and under a natural assumption about the dimensionality of the system under test, we present an experimental proof of the non-classicality of a DCE based on the violation of a dimension witness inequality. Our conclusion is reached in a device-independent and loophole-free manner, that is, based solely on the observed data and without the need on any assumptions about the measurement apparatus.
\end{abstract}

\maketitle
\textit{Introduction ---}  The wave or particle nature of light is amongst the oldest debates in physics \cite{Sabra1981}. The famous double slit experiment demonstrating interference \cite{Young1807} settled the question for a while. However, with the establishment of quantum mechanics, the conundrum was back again, giving rise to one of the most counterintuitive features of quantum theory, the wave-particle duality \cite{bohr,englert}, i.e. depending on the experimental apparatus, a quantum system can exhibit a particle or wave behavior. For instance, a photon in a Mach-Zehnder interferometer displays interference (wave-like) or no-interference (particle-like) depending whether a second beam splitter is put at the intersection of the two interferometric arms. As argued by Bohr \cite{bohr}, however, one cannot assign to each quantum system a definite and immutable label that determines its character. Wave and particle are complementary concepts, with neither excluding the other.

In order to rule out an interpretation in which a quantum system is intrinsically \textit{either} a wave \textit{or} a particle,
Wheeler proposed his famous delayed choice experiment (DCE) \cite{Wheeler1,Wheeler2}, which requires the experimental arrangement revealing or not the interference pattern, to be decided \emph{after} the photon has entered the
interferometer. Over the years, the DCE idea has been explored in several directions \cite{Ma}, including implementations on different physical platforms \cite{Roy,jacques}. More recently, interest in the topic has been revived by the proposal of a quantum DCE \cite{Ioniciou,Accaise,Tang,shalbo}, where the presence or absence of the second beam splitter is decided by a photon in a quantum superposition. Following that, wave-particle duality has also been related to quantum entanglement \cite{Adil,kaiser,Peruzzo,Ioniciou2}. Excluded the possibility of retro-causal influences, where the late choice might change the nature of the photon back in time, both Wheeler's and the quantum DCE prove the incompatibility with the quantum predictions of any model whereby the photon has a definite intrinsic wave or particle nature. Nonetheless, recent results based on causal modeling \cite{Pearl2009} have shown that a DCE can be seen as a particular instance of a prepare and measure (PAM) scenario \cite{pam1,Bowles}, also revealing that a classical hidden variable (HV) model can in fact yield the results of the DCE statistics \cite{rossi2017,Rafael18}. The causal approach offers a way to circumvent debatable definitions and probe the non-classical nature of an experiment based solely on the empirical data at hand and some mild causal assumptions. This kind of approach is commonly referred to as \textit{device-independent} (DI) and provides tools with a vast range of applications \cite{Mayers2004,Barrett2005,Acin2006,Brunner2008,Buhrman2010,Pironio2010,Chaves2015,Lunghi2015,Instrumental}.

In spite of the fact that Wheeler's and quantum DCEs are of classical nature from the DI perspective, it has also been shown that minor modifications of the original proposals can rule out, with a DI protocol, HV models as a possible explanation  \cite{Rafael18}. In this new framework, apart from the assumption of non-retrocausality, the HV is also assumed to have the same dimension as the quantum system it is supposed to mimic. Thus, the non-classicality in the DCE can be tested via well developed dimensional witness inequalities \cite{pam1,Bowles} that, if violated, prove that any classical explanation requires physical systems with higher dimension than its quantum counterpart, or else has to resort to retro-causal influences.

\begin{figure}[htb]
\includegraphics[width=1\columnwidth]{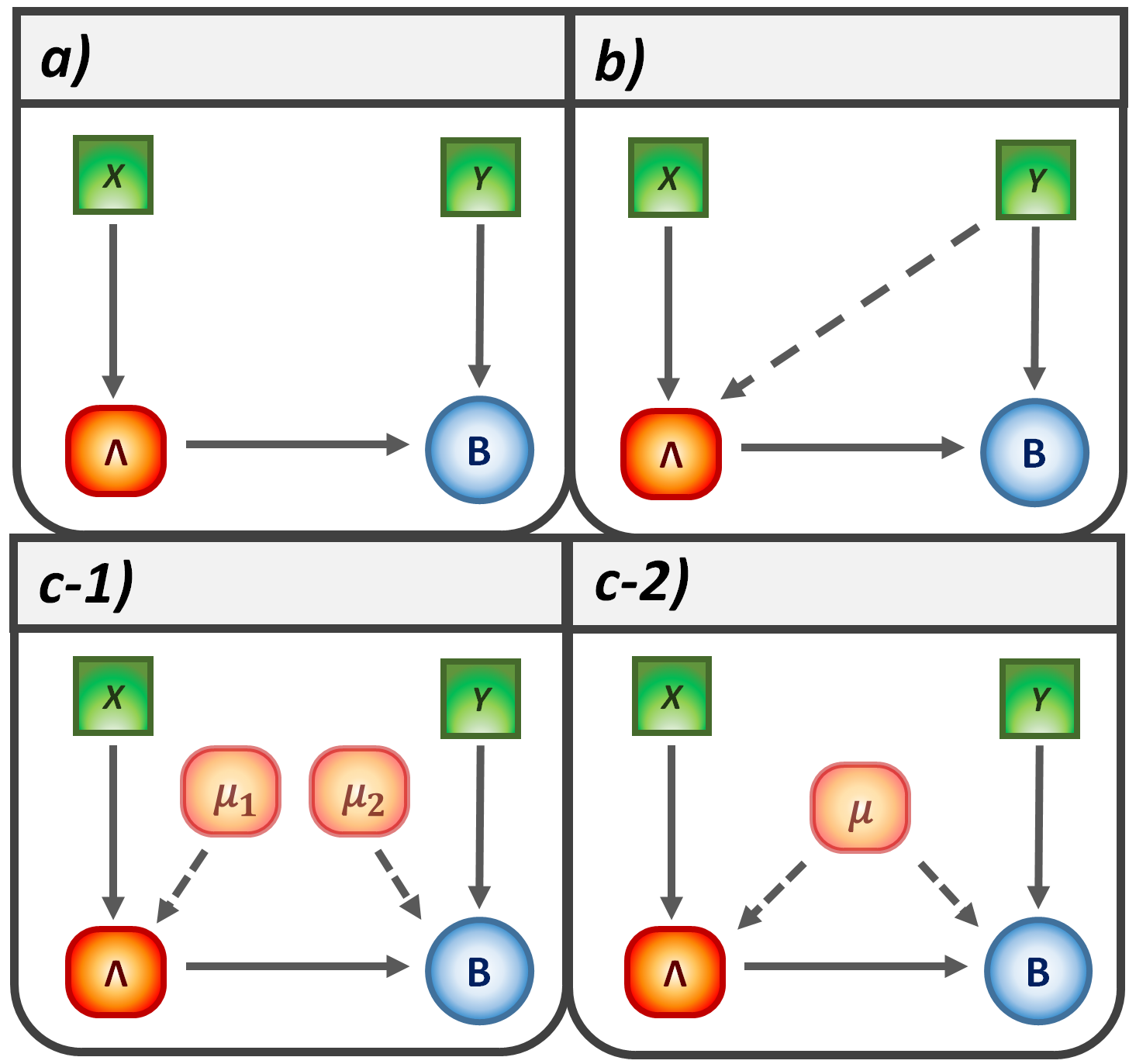}
\caption{\textbf{Causal structures representing the delayed choice experiment}
\textbf{a)} \textit{Prepare and Measure} (PAM) scenario, where a classical variable $\Lambda$, or a quantum state $\rho$ depending on whether we refer to classical or quantum description, depends on a classical variable $X$ while the choice of which measurement to perform is denoted by $Y$. Under the assumption of non-retrocausality, the variables $Y$ and $\Lambda$ should be statistically independent, that is, $p(\lambda,y)=p(\lambda)p(y)$. \textbf{b)} Relaxation on the assumption of no retrocausality in a PAM scenario: to explain the violation of the dimensional witnesses reported through a classical HV model, we need to allow retro-causality. \textbf{c)} PAM scenario affected by noise terms. \textbf{c-1)} The preparation and measurement devices are uncorrelated, since they are affected by two independent noise terms $\mu_1$ and $\mu_2$. \textbf{c-2)} Preparation and measurement stages are allowed to have pre-established correlations mediated by the variable $\mu$.}
\label{fig:CausalStructures}
\end{figure}

In this letter, we report on the realization of device-independent tests of non-classicality in two complementary photonic quantum DCE experiments. First, under the reasonable assumption that the preparation and measurement stages in our experiment are uncorrelated (see Fig. \ref{fig:CausalStructures}c-1), we provide an experimental detection loophole-free violation of a dimensional witness.
Second, we provide more resources to the potential HV model, allowing for correlations between the preparation and measurement devices, i.e. relaxing the hypothesis of independent noise terms affecting each variable  (see Fig. \ref{fig:CausalStructures}c-2). Even in this case, under the fair-sampling assumption, we can violate a dimensional witness inequality. Altogether, our results provide the first DI verification of the true quantum nature of delayed choice experiment, without the need to resort to quantum entanglement \cite{Adil,kaiser,Peruzzo}.

\begin{figure}[b]
\includegraphics[width=0.5\textwidth]{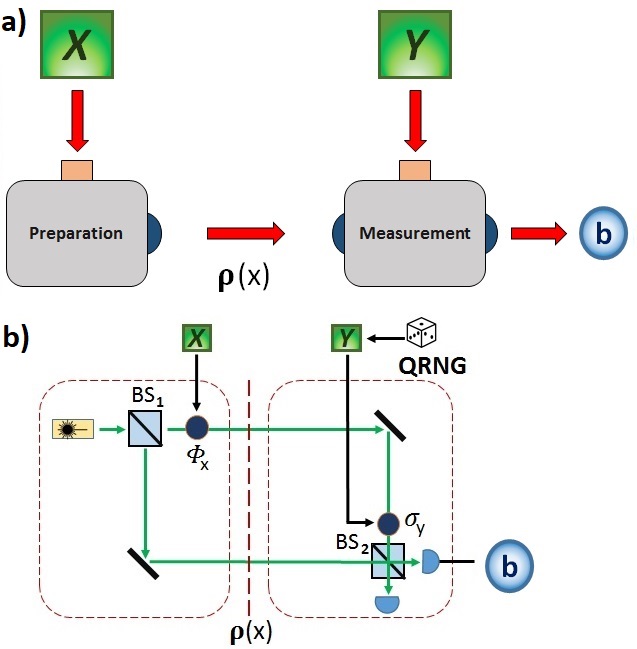}
\caption{\textbf{Delayed choice experiment as a PAM  scenario.} \textbf{a)} In a PAM scenario, an input $x$ selects a quantum state $\rho_{x}$, which is then injected in a measurement station. The other input $y$ selects the observable and b is its output. \textbf{b)} Possible implementation of the modified DCE as a PAM scenario, through a Mach-Zehnder interferometer. X and Y select respectively among three and two phases, that are added on one arm of the interferometer. The output b is 1 when the highlighted detector clicks and 0 when it does not.}
        \label{fig:PAM}
\end{figure}

\begin{figure*}[ht!]
\includegraphics[width=1\textwidth]{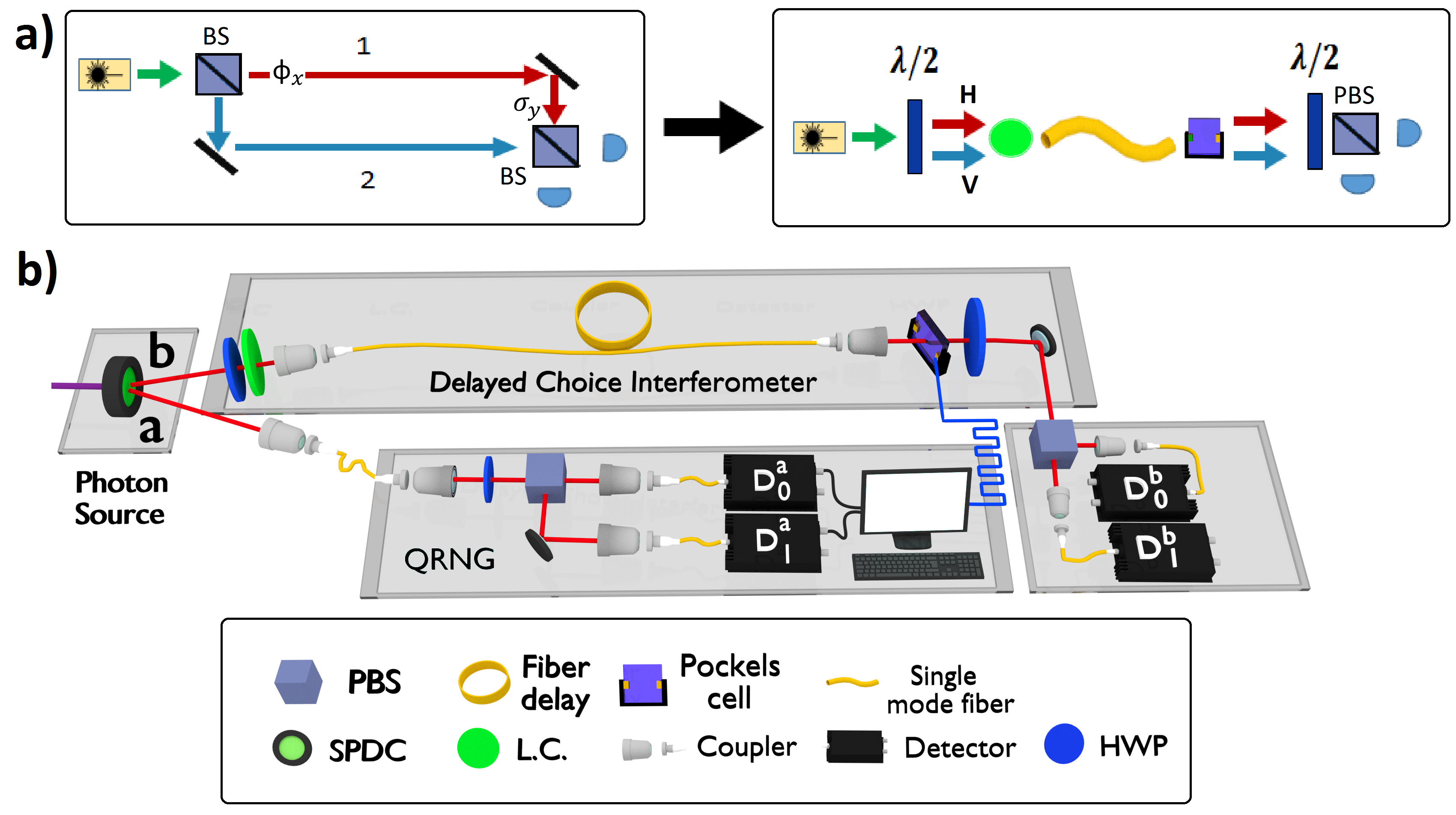}
\caption{\textbf{Experimental apparatus for the violation of the dimensional witness.} \textbf{a)} A pictorial representation of the modified DCE proposed in \cite{Rafael18} (left) and our apparatus, exploiting the equivalent, in polarization, of a Mach-Zehnder interferometer (right). \textbf{b)} Two photons are generated via spontaneous parametric down-conversion process in a ppKTP non-linear crystal. One photon is the trigger and is sent to the Quantum Random Number Generation (QRNG) stage (path a) to select the phase shift $\sigma_{y}$, between two values $\sigma_{0}$ and $\sigma_{1}$. When $D^{a}_{0}$ clicks, it triggers the application of a 1,210 V voltage on the Pockels cell on path b, which adds the required phase shift $\sigma_{1}$, otherwise $\sigma_{y}$=$\sigma_{0}$=0. Since the photon is horizontally polarized, the action of a rotated HWP before the PBS ensures an intrinsically random and uniformly distributed choice. The second photon (vertically polarized) enters the polarization Mach-Zehnder interferometer, composed by a 22.5$^{\circ}$ rotated HWP, to generate the superposition of the entering polarization and its orthogonal, a liquid crystal (LC), which introduces a phase shift $\phi_{x}$, between the horizontal and vertical polarization components. Then, the photon is delayed of 710 ns, employing a 140 m long single-mode fiber. After leaving the fiber, the photon passes through the Pockels cell, which, when triggered, introduces the $\pi/ 2$ phase shift. Just after the Pockels cell, another 22.5$^{\circ}$ rotated HWP closes the polarization Mach-Zehnder interferometer. A PBS performs the required projective measurement.}
\label{fig:EXP}
\end{figure*}

\textit{A delayed choice experiment as a prepare and measure scenario ---} As recently noted in \cite{Rafael18}, a DCE (Fig. \ref{fig:PAM}b) can be seen as a particular instance of a PAM scenario (Fig. \ref{fig:PAM}a). Upon receiving an input $x$, a state preparation device emits a quantum state $\rho_{x}$ (Fig. \ref{fig:PAM}a). This state is then sent to a measurement device, where the measurement to be performed is selected by another input $y$, producing output $b$. The quantum description of the experiment produces probability distributions according to the Born rule
\begin{equation}
P_{\mathrm{Q}}(b \vert x,y)= \mathrm{Tr}\left(\rho_{x} M_{y}^{b}\right),
\end{equation}
where $M_{y}^{b}$ is a positive operator such that $\sum_{b}M_{y}^{b}=\openone$ for all $y$. In turn, in a classical description modeled by a HV causal model, the probability can be decomposed as
\begin{equation}
\label{HVmodel}
P_{\mathrm{C}}(b\vert x,y)= \sum_{\lambda}p(\lambda \vert x)p(b\vert \lambda, y),
\end{equation}
where the classical variable $\lambda$ can also be interpreted as a quantum state diagonal in a given fixed basis. The central causal assumption in a PAM scenario as well as in a DCE is the fact that the choice $Y$ does not have causal influence over the preparation stage of the state $\rho_x$ (or $\Lambda$ in the HV description, see Fig. \ref{fig:CausalStructures}). If there are no bounds on the Hilbert space dimension or on the cardinality of $\lambda$, any statistics can be obtained and the classical and quantum models give the same predictions. However, if the dimension is limited, there are quantum distributions that can only be generated by a classical model with higher dimensionality \cite{pam1,Bowles,Rafael18}.

\begin{figure}[h!]
\includegraphics[width=1\columnwidth]{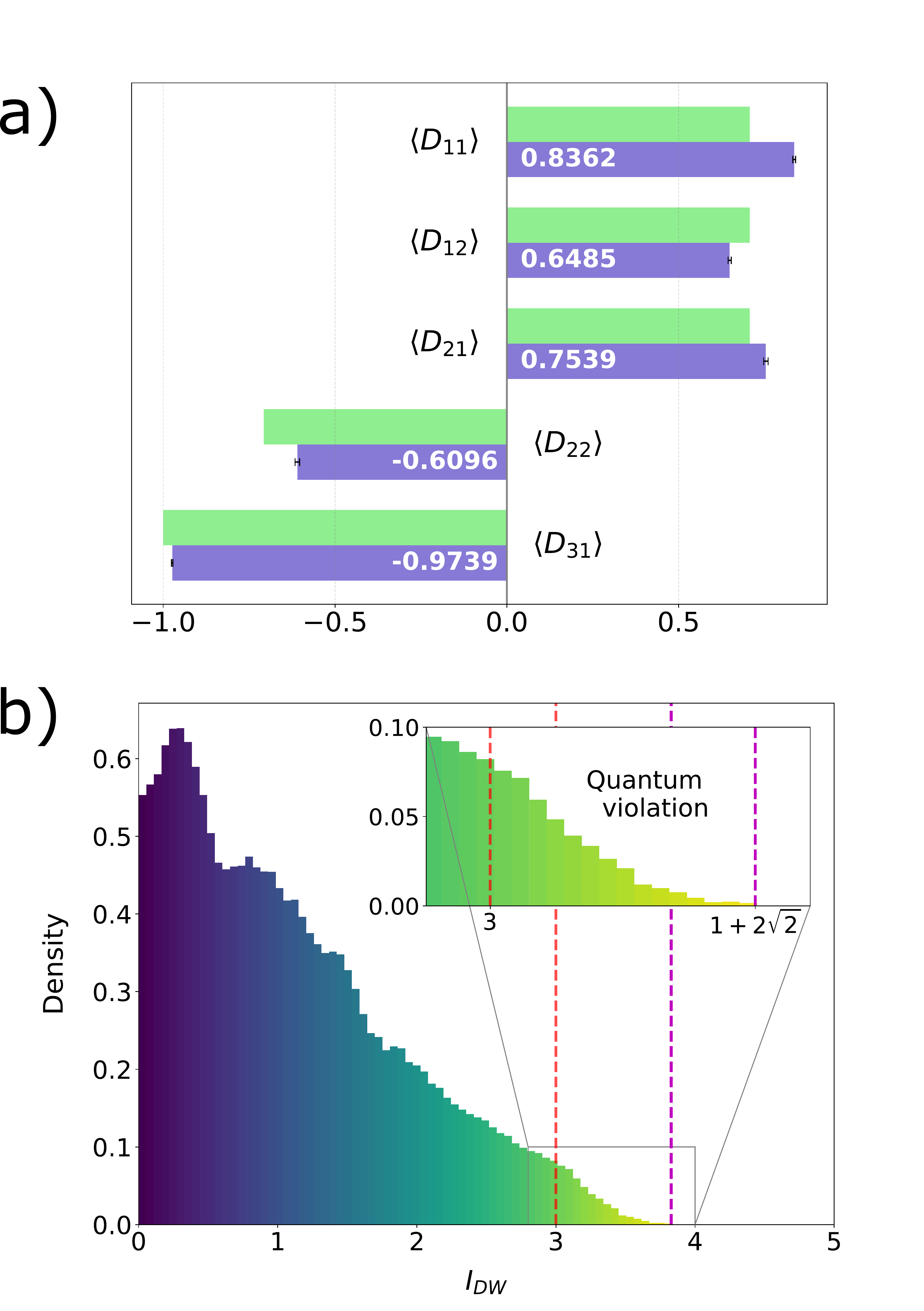}
\caption{\textbf{Experimental violation of dimensional witness $I_{DW}$. a)} Experimental values (purple bars) of each of the terms in Eq. \ref{eq:IDW}, for the case of maximal violation: $\braket{B_{11}}$, $\braket{B_{12}}$, $\braket{B_{21}}$, $\braket{B_{22}}$, $\braket{B_{31}}$, along with their uncertainties, estimated by the propagation of the Poissonian uncertainty of counts. The corresponding theoretical values (green bars) for the optimal tuple of $\phi_{x}$ ($7/4 \pi$, $5/4 \pi$, $\pi/2$), are respectively ($1/\sqrt{2}$, $1/\sqrt{2}$, $1/\sqrt{2}$, -$1/\sqrt{2}$ and -1) \textbf{b)} Histogram of the experimental probability distribution to obtain each value of $I_{DW}$, from 0 (lowest value) to 5 (post-quantum bound). The dashed pink line indicates the classical limit of $I_{DW}$, 3, while the purple one indicates the quantum limit, 1+2$\sqrt{2}\approx3.828$. Those probabilities are evaluated as the frequencies of the obtained value of $I_{DW}$ over $70^{3}$ tuples of phase shifts $\phi_{x}$. Some of these tuples present a violation of the classical limit, as can be seen in the enlarged region of the histogram.}
\label{fig:histIDW}
\end{figure}

The application of this approach to Wheeler's and the quantum DCE --where the measurement choices correspond to the presence or absence of a second beam splitter -- show that the HV model \eqref{HVmodel} with dimension equal to $d=2$ can reproduce the quantum predictions \cite{Rafael18}. Indeed, as expected, since the DCE involves a single photon in two modes, encoding at most a single classical bit \cite{Holevo1973}, $\Lambda$ should likewise be binary. However, if instead the second beam splitter is always present and the measurement choice now corresponds to an extra phase $\sigma_{y}$ (imprinted long after the photon has entered the interferometer), 2-dimensional HV models can no longer reproduce the quantum correlations.

As shown in Fig. \ref{fig:EXP}, we have implemented the DCE/PAM causal structure using a photonic platform, which is a variation of the apparatus shown in Fig. \ref{fig:PAM}b, involving the equivalent in polarization of a Mach-Zehnder interferometer. 
The interferometer is composed by a first and second half-wave plates (HWP). The first HWP acts as the initial beam splitter of a  Mach-Zehnder generating a superposition of the two orthogonal polarizations mimicking the spatial paths, while the second takes the place of the final beam splitter, recombining the polarizations and allowing interference between them. In the preparation stage, after the first half-wave plate, the variable $X$ randomly selects a phase $\phi_{x}$ inserted between the two polarizations via a liquid crystal. In the measurement stage, to ensure the \textit{delayed choice}, we employ two key elements: i) a quantum random number Generator (QRNG) based on a measurement performed on a single photon, which randomly impinges the extra phase $\sigma_{y}$, ii) a fast electro-optics modulator (Pockels cell) implementing this phase within the Mach-Zehnder interferometer. The measurement is realized by a polarizing beam splitter (PBS) whose outputs are sent to two detectors. With this setup we have violated two distinct types of dimension witness inequalities. In both cases, we need two measurement setups, one where an extra phase $\sigma_{1}=\pi/2$ is introduced and a second with no extra phase $\sigma_{2}=0$.

The QRNG is implemented through a horizontally polarized photon (indicated as the trigger), sent on path ``a''. Its polarization is rotated by a HWP to generate $\ket{+}=(\ket{H}+\ket{V})/2$ and then sent through a PBS. When the emerging photon is horizontally (vertically) polarized, the selected phase shift is $\sigma_{1}$ ($\sigma_{2}$). This procedure gives the same probability to both $\sigma_{1}$ and $\sigma_{2}$, ensuring an intrinsically random choice. The trigger photon is generated at the same time of the photon which enters the Mach-Zehnder (indicated as the signal), and thus can guarantee that the measurement choice is performed after the signal photon has entered the interferometer. The Pockels cell allows the choice between $\sigma_{1}$ and $\sigma_{2}$ to be made swiftly, since its response time is within the order of nanoseconds. When the detector $D^{a}_{0}$ clicks, the electronic signal of the avalanche photo-diode is split, so it can both be registered by the coincidence counter and trigger the Pockels cell that introduces a $\pi/2$ phase shift between the two superimposed polarizations. When $D^{a}_{1}$ clicks, no voltage is applied introducing no extra phase. To allow this active feed-forward from path ``a'' to path ``b'', the latter is stretched, through a 140 m long single-mode optical fiber, in order to achieve a delay of $\sim$710 ns.

First, we have tested the $W_{2}$ dimensional witness allowing to rule out in DI manner any HV model where the preparation and measurement devices are independent (see Fig. \ref{fig:CausalStructures}c-1). Its experimental value is obtained selecting four possible $x$ preparations and two possible measurement $y$, as the determinant of the following 2x2 matrix:
\begin{equation}
W_{2} = 
\begin{pmatrix}
    & p(1,1)-p(2,1) & p(3,1)-p(4,1)  \\
    & p(1,2)-p(2,2) & p(3,2)-p(4,2)   \\
\end{pmatrix}
\end{equation}
where $p(i,j)= p(d=0|x=i, y=j)$, with $i=1,..., 4$ and $j=1, 2$. Any 2-dimensional classical system satisfies $|{\rm det}(W_{2})|=0$. In our setup, selecting the following values for $\phi_{x}$ and $\sigma_{y}$: $\phi_{1}=0$, $\phi_{2}=\pi$, $\phi_{3}= -\pi/2$, $\phi_{4}= \pi/2$  and $\sigma_{1}=\pi/2$, $\sigma_{2}=0$, the quantum probabilities achieve $W_{2}=1$, thus violating the classical prediction. Experimentally, we obtained $W_{2}= 0.951 \pm 0.010$.
This result was obtained by post-selecting only those events for which there is a simultaneous arrival of a signal photon, on path ``b'', 
and a trigger photon, on path ``a''. This is equivalent to having detectors with the highest efficiency possible, $\eta=1$, and ideal transmittances $T_a, T_b=1$ on the interferometric arms, since eventual lost photons are not considered. Under this fair-sampling assumption, we say that $b=0$ when we have a coincidence between the trigger photon and $D^{b}_{0}$, and $b=1$ when the coincidence is with $D^{b}_{1}$. However, $W_{2}$ can also be violated in a loophole-free manner without the need of any further assumptions. To this aim, we consider each click of the trigger photon as an experiment run and assign to $b$ the value 1 when a coincidence is registered with $D^{b}_{0}$ and $0$ both when the coincidence is with $D^{b}_{1}$ and when the photon is lost. In other words, without post-selecting on the coincidence counts, whenever the trigger clicks, we always assign a value to $b$, even if the photon on path ``b'' is lost.
In this case, the real efficiency ($\eta  \approx 0.025$) of
$D^{b}_{0}$ and eventual imperfections in the transmittances affect the witness value, that is now given by $|det(W_{2}^{\eta, T_{a}, T_{b}})|=
\eta^{2}T_{a}^{2} T_{b}^{2}|det(W_{2}^{ \eta=1, T_{a}, T_{b} =1})|$ and thus in principle can be violated any $\eta, T_{a}, T_{b} > 0$. Indeed, we obtain a violation given by $W_{2}=(1.23 \pm 0.03) 10^{-4}$ and demonstrated that the obtained statistics cannot be explained by a classical HV model even in the presence of major experimental imperfections.

As a complementary experiment we have also tested stronger HV models where now the preparation and measurement devices can be correlated (see Fig. \ref{fig:CausalStructures}c-2). Within a scenario with three possible $x$ inputs, while $y$ is still dichotomic we consider the following dimension witness \cite{pam1}:
\begin{equation}
I_{DW}=|\braket{B_{11}}+\braket{B_{12}}+\braket{B_{21}}-\braket{B_{22}}-\braket{B_{31}}|
\label{eq:IDW}
\end{equation}
where $\braket{B_{xy}} = p(b=0|x,y)-p(b=1|x,y)$, with $x=1,2,3$ and $y=1,2$.\\
The classical bound for $I_{DW}$ is 3, while the quantum bound is $Q_{2}=1+2\sqrt{2}=3.828$ \cite{pam1, pam2,Rafael18}. A value of $I_{DW}$ higher than the classical bound thus rules out any 2-dimensional HV model that does not allow for retro-causality, even with no special assumptions on the potential pre-established correlations.

We experimentally evaluated $70^{3}$ values of $I_{DW}$, each one corresponding to a different tuple of $\phi_{x}$, always with $\sigma_1=\pi/2$ and  $\sigma_2=0$. As it is shown in Fig. \ref{fig:histIDW}b, for some configurations, we are able to violate the classical bound and obtain values within the range between 3 and 1+2$\sqrt{2}$. Specifically, in blue, we plot the experimental probability distribution to obtain each $I_{DW}$ value and highlight the classical (pink) and quantum (purple) upper bounds. This procedure, where we span different values of $\phi_{x}$, instead of selecting \textit{a priori} the theoretical phase shifts which maximize the violation, highlights the device-independent nature of our test, since it does not rely on any trust on the experimental apparatus to insert exactly a $\phi_{x}$ phase shift between the two interferometric arms.

The highest experimental $I_{DW}$ value is $3.822 \pm 0.011$, as reported in Fig. \ref{fig:histIDW}a, through the experimental values of each term in Eq.  \ref{eq:IDW}. Specifically, the choice of $\phi_{x}$ and $\sigma_{y}$ which maximize $I_{DW}$ is the following: $\phi_{1}=7/4 \pi$, $\phi_{2}=5/4 \pi$, $\phi_{3}= \pi/2$  and $\sigma_{1}=\pi/2$, $\sigma_{2}=0$. In order to explain this result within a classical framework, we would need to relax the assumption of complete absence of influence between Y and $\Lambda$, that is, to allow retro-causality. As detailed in \cite{Rafael18}, the minimum amount of retro-causality as quantified by a measure $R_{Y \to \Lambda}$ is given by
\begin{equation}
min (R_{Y \to \Lambda}) = max \Biggl[ \frac{I_{DW}-3}{4}, 0 \Biggr]. 
\end{equation}
Thus, our experiment indeed falsifies even stronger models where retro-causality is allowed but below the threshold $R_{Y \to \Lambda}=0.2055 \pm 0.0027$.

\textit{Discussion ---} In this work we realized for the first time a modified DCE \cite{Rafael18}, that allows to discriminate, in a device independent way and without using entanglement, quantum predictions from those of an non retro-causal HV theory. The only crucial assumption is that the dimension of the HV is dichotomic (e.g., assuming "wave" or "particle" values), that is, the classical model has the same dimension as the quantum system under test. Our experiment is based on an optical interferometer employing polarization degree of freedom of photons to probe their non-classical nature. In turn, the delayed choice of the observables is achieved by a quantum random generator and an active feed-forward which triggers a fast electro-optical modulator. With this setup, we measured two DI dimensional witnesses: i) $W_2$, whose experimental value overcomes the classical expected value zero by 41 standard deviations and ii) $I_{DW}$, that in a classical two-dimensional framework should be lower than 3 and whose experimental value violates the inequality by more than 70 standard deviations, being compatible with the quantum prediction of $\approx 3.828$.
The violation of $W_2$ is achieved in a detection-loophole free manner since it is tolerant against arbitrary losses and efficiencies but requires the assumption of independent preparation and measurement devices, an assumption that can be withdrawn in the violation of $I_{DW}$ (requiring the fair-sampling assumption).
Our results highlight the relevance of revisiting foundational experiments from a causal perspective and we expect it might trigger further applications of causal modeling to other fundamental tests as well as to applications such as randomness generation \cite{Lunghi2015} or certification \cite{Li2012}.

\emph{Note added}.
 The present experiment and a similar modified delayed choice experiment, of H.-L. Huang  et al., were carried out simultaneously and independently.

\bigskip

\begin{acknowledgments}
\textbf{Acknowledgements.}
This work was supported by the QuantERA project HIPHOP (project ID 731473), the Sapienza University funding Progetto d'Ateneo, the Brazilian ministries MEC and MCTIC and agency CNPq.
\end{acknowledgments}

\end{document}